\documentstyle[twoside]{article}

\catcode`\@=11
\long\def\@makefntext#1{
\protect\noindent \hbox to 3.2pt {\hskip-.9pt  
$^{{\eightrm\@thefnmark}}$\hfil}#1\hfill}		%CAN BE USED 

\def\@makefnmark{\hbox to 0pt{$^{\@thefnmark}$\hss}}	%ORIGINAL 
	
\def\ps@myheadings{\let\@mkboth\@gobbletwo
\def\@oddhead{\hbox{}
\rightmark\hfil\eightrm\thepage}   
\def\@oddfoot{}\def\@evenhead{\eightrm\thepage\hfil
\leftmark\hbox{}}\def\@evenfoot{}
\def\sectionmark##1{}\def\subsectionmark##1{}}

\oddsidemargin=\evensidemargin
\newcounter{sectionc}\newcounter{subsectionc}\newcounter{subsubsectionc}
\renewcommand{\section}[1] {\vspace{12pt}\addtocounter{sectionc}{1} 
\setcounter{subsectionc}{0}\setcounter{subsubsectionc}{0}\noindent 
	{\tenbf\thesectionc. #1}\par\vspace{5pt}}
\renewcommand{\subsection}[1] {\vspace{12pt}\addtocounter{subsectionc}{1} 
	\setcounter{subsubsectionc}{0}\noindent 
	{\bf\thesectionc.\thesubsectionc. {\kern1pt \bfit #1}}\par\vspace{5pt}}
\renewcommand{\subsubsection}[1] {\vspace{12pt}\addtocounter{subsubsectionc}{1}
	\noindent{\tenrm\thesectionc.\thesubsectionc.\thesubsubsectionc.
	{\kern1pt \tenit #1}}\par\vspace{5pt}}
\newcommand{\nonumsection}[1] {\vspace{12pt}\noindent{\tenbf #1}
	\par\vspace{5pt}}

\newcounter{appendixc}
\newcounter{subappendixc}[appendixc]
\newcounter{subsubappendixc}[subappendixc]
\renewcommand{\thesubappendixc}{\Alph{appendixc}.\arabic{subappendixc}}
\renewcommand{\thesubsubappendixc}
	{\Alph{appendixc}.\arabic{subappendixc}.\arabic{subsubappendixc}}

\renewcommand{\appendix}[1] {\vspace{12pt}
        \refstepcounter{appendixc}
        \setcounter{figure}{0}
        \setcounter{table}{0}
        \setcounter{lemma}{0}
        \setcounter{theorem}{0}
        \setcounter{corollary}{0}
        \setcounter{definition}{0}
        \setcounter{equation}{0}
        \renewcommand{\thefigure}{\Alph{appendixc}.\arabic{figure}}
        \renewcommand{\thetable}{\Alph{appendixc}.\arabic{table}}
        \renewcommand{\theappendixc}{\Alph{appendixc}}
        \renewcommand{\thelemma}{\Alph{appendixc}.\arabic{lemma}}
        \renewcommand{\thetheorem}{\Alph{appendixc}.\arabic{theorem}}
        \renewcommand{\thedefinition}{\Alph{appendixc}.\arabic{definition}}
        \renewcommand{\thecorollary}{\Alph{appendixc}.\arabic{corollary}}
        \renewcommand{\theequation}{\Alph{appendixc}.\arabic{equation}}
        \noindent{\tenbf Appendix \theappendixc #1}\par\vspace{5pt}}
\newcommand{\subappendix}[1] {\vspace{12pt}
        \refstepcounter{subappendixc}
        \noindent{\bf Appendix \thesubappendixc. {\kern1pt \bfit #1}}
	\par\vspace{5pt}}
\newcommand{\subsubappendix}[1] {\vspace{12pt}
        \refstepcounter{subsubappendixc}
        \noindent{\rm Appendix \thesubsubappendixc. {\kern1pt \tenit #1}}
	\par\vspace{5pt}}

\newcommand{\textlineskip}{\baselineskip=13pt}
\newcommand{\smalllineskip}{\baselineskip=10pt}

\def\eightcirc{
\begin{picture}(0,0)
\put(4.4,1.8){\circle{6.5}}
\end{picture}}
\def\eightcopyright{\eightcirc\kern2.7pt\hbox{\eightrm c}} 

\newcommand{\copyrightheading}[1]
	{\vspace*{-2.5cm}\smalllineskip{\flushleft
	{\footnotesize International Journal of Modern Physics A, #1}\\
	{\footnotesize $\eightcopyright$\, World Scientific Publishing
	 Company}\\
	 }}

\def\abstracts#1#2#3{{
	\centering{\begin{minipage}{4.5in}\baselineskip=10pt\footnotesize
	\parindent=0pt #1\par 
	\parindent=15pt #2\par
	\parindent=15pt #3
	\end{minipage}}\par}} 

\renewenvironment{thebibliography}[1]
	{\frenchspacing
	 \ninerm\baselineskip=11pt
	 \begin{list}{\arabic{enumi}.}
	{\usecounter{enumi}\setlength{\parsep}{0pt}
	 \setlength{\leftmargin 12.7pt}{\rightmargin 0pt} %FOR 1--9 ITEMS
	 \setlength{\itemsep}{0pt} \settowidth
	{\labelwidth}{#1.}\sloppy}}{\end{list}}

\newcounter{itemlistc}
\newcounter{romanlistc}
\newcounter{alphlistc}
\newcounter{arabiclistc}

\newcommand{\fcaption}[1]{
        \refstepcounter{figure}
        \setbox\@tempboxa = \hbox{\footnotesize Fig.~\thefigure. #1}
        \ifdim \wd\@tempboxa > 5in
           {\begin{center}
        \parbox{5in}{\footnotesize\smalllineskip Fig.~\thefigure. #1}
            \end{center}}
        \else
             {\begin{center}
             {\footnotesize Fig.~\thefigure. #1}
              \end{center}}
        \fi}

\newcommand{\tcaption}[1]{
        \refstepcounter{table}
        \setbox\@tempboxa = \hbox{\footnotesize Table~\thetable. #1}
        \ifdim \wd\@tempboxa > 5in
           {\begin{center}
        \parbox{5in}{\footnotesize\smalllineskip Table~\thetable. #1}
            \end{center}}
        \else
             {\begin{center}
             {\footnotesize Table~\thetable. #1}
              \end{center}}
        \fi}

\def\@citex[#1]#2{\if@filesw\immediate\write\@auxout
	{\string\citation{#2}}\fi
\def\@citea{}\@cite{\@for\@citeb:=#2\do
	{\@citea\def\@citea{,}\@ifundefined
	{b@\@citeb}{{\bf ?}\@warning
	{Citation `\@citeb' on page \thepage \space undefined}}
	{\csname b@\@citeb\endcsname}}}{#1}}

\newif\if@cghi
\def\cite{\@cghitrue\@ifnextchar [{\@tempswatrue
	\@citex}{\@tempswafalse\@citex[]}}
\def\citelow{\@cghifalse\@ifnextchar [{\@tempswatrue
	\@citex}{\@tempswafalse\@citex[]}}
\def\@cite#1#2{{$\null^{#1}$\if@tempswa\typeout
	{IJCGA warning: optional citation argument 
	ignored: `#2'} \fi}}

\def\pmb#1{\setbox0=\hbox{#1}
	\kern-.025em\copy0\kern-\wd0
	\kern.05em\copy0\kern-\wd0
	\kern-.025em\raise.0433em\box0}

\def\fnt#1#2{\footnotetext{\kern-.3em
	{$^{\mbox{\scriptsize #1}}$}{#2}}}

\def\fpage#1{\begingroup
\voffset=.3in
\thispagestyle{empty}\begin{table}[b]\centerline{\footnotesize #1}
	\end{table}\endgroup}

\def\runninghead#1#2{\pagestyle{myheadings}
\markboth{{\protect\footnotesize\it{\quad #1}}\hfill}
{\hfill{\protect\footnotesize\it{#2\quad}}}}
\font\tenrm=cmr10
\font\tenit=cmti10 
\font\tenbf=cmbx10
\font\bfit=cmbxti10 at 10pt
\font\ninerm=cmr9

\font\eightrm=cmr8

\def\qed{\hbox{${\vcenter{\vbox{			%HOLLOW SQUARE
   \hrule height 0.4pt\hbox{\vrule width 0.4pt height 6pt
   \kern5pt\vrule width 0.4pt}\hrule height 0.4pt}}}$}}

\begin{document}

\runninghead{Random Values of the Physical Parameters} {Random Values of
the Physical Parameters}

\normalsize\textlineskip
\thispagestyle{empty}
\setcounter{page}{1}

\copyrightheading{}			%{Vol. 0, No. 0 (1993) 000--000}

\vspace*{0.88truein}

\fpage{1}
\centerline{\bf RANDOM VALUES OF THE PHYSICAL PARAMETERS}
\vspace*{0.37truein}
\centerline{\footnotesize JOHN F. DONOGHUE}
\vspace*{0.015truein}
\centerline{\footnotesize\it Physics Department, University
of Massachusetts, Amherst MA 01003}
%\vspace*{0.225truein}
%\publisher{(received date)}{(revised date)}

\vspace*{0.21truein}
\abstracts{I briefly describe two motivations, two mechanisms and 
two possible tests of the hypothesis that the physical parameters
of the ground state of a theory can vary in different 
regions of the universe.}{}{}

%\textlineskip			%) USE THIS MEASUREMENT WHEN THERE IS
%\vspace*{12pt}			%) NO SECTION HEADING

\vspace*{1pt}\textlineskip	%) USE THIS MEASUREMENT WHEN THERE IS
\section{Introduction}	%) A SECTION HEADING
\vspace*{-0.5pt}
\noindent

The Standard Model has a unique ground state
(modulo SU(2)$_L$ rotations) and the 19 parameters of the
model are uniquely fixed. However, when considering theories beyond
the Standard Model, we have no evidence that this uniqueness will be
preserved. In particular, string theory has continuous families
of ground states, each with different values of the physical
parameters, and we presently do not know the mechanism that
distinguishes among them. Is the unique ground state of
string theory that one with SU(3) X SU(2) X U(1) and $m_u=3$~MeV,
$m_d=6$~MeV etc.? An alternative is that many, perhaps all,
of the ground states can be realized in different regions of the universe,
depending in some way on the past history of that region.
At present we have no reason to rule out this possibility.
However, if this is the case, we need to approach the theory
in fundamentally different ways. This talk is devoted to
some preliminary work in the context of random physical
parameters$^1$.

\section{Two Motivations}
\noindent
Theories with variable parameters must       
involve anthropic
constraints. In the space of all possible parameters, most domains
involve combinations of parameters which are not suitable for life of any form.
While many physicists have a negative reaction to 
anthropic ideas, they are a natural
outcome of having domains with different parameters in the universe.

The strongest motivation for variable parameters and
anthropic constraints comes from the cosmological constant 
$\Lambda$.
The natural scale of this parameter is so large that the
anthropically allowed region is an extremely
tiny portion of parameter space. Most other values of $\Lambda$ 
would not
allow matter to clump into galaxies, or would make the 
universe extremely
short lived. Unless we are able to uncover a generic mechanism to 
produce a small non-zero value of $\Lambda$,
it would be remarkably lucky if the unique ground
state of the ultimate theory just happened to fall in the
anthropically allowed range. It seems more plausible that in
a large ensemble of domains with various parameters, a few 
of those domains have parameters which happened
to fall in the allowed range. We could only find ourselves in
such a domain.

A second motivation concerns the Higgs vacuum expectation value.
If the Higgs vev was much larger (with the other parameters fixed)
none of the complex elements would exist since the mass differences
of quarks would exceed the 10 MeV per nucleon binding energy of nuclei,
allowing decay of all quarks down to the lightest quark. A world of
hydrogen alone likely does not have the complexity needed for life.
This suggests that an anthropic constraint may also be at work
to require that the weak scale be close to the QCD scale.

Note that anthropic constraints do not ``solve'' the
cosmological constant or the Higgs vev problems Instead they
suggest the possibility that these are accidents of history, not
really problems to be solved. The real function of these considerations
is to motivate a search for theories with variable parameters.

\section{Two Mechanisms}
\noindent
Scalar fields can be frozen at random values by the expansion
of the universe if their potential is flat enough. This mechanism is 
used in inflationary theories to temporarily keep the scalar field from
rolling down the potential. However, if the potential is yet 
flatter, compared to the Hubble expansion, the scalar field will
be frozen longer$^{1,2,3}$.
If in the early history of the universe these scalar fields were
initially fluctuating, they could get frozen at different values
in different regions, influencing the cosmological constant in each region.
For the present
Hubble expansion $H_o = 10^{-122} M_P^2$, to freeze the scalar field
requires a {\it{very}} flat potential. 
From this extreme flatness we can conclude
that matter fields cannot couple to the scalar. (Otherwise loops of
these other fields would generate a potential which  would be too large
unless there was extreme fine-tuning.) This implies that only 
the cosmological constant would be variable with this mechanism.

A more exotic mechanism involves four-form field strengths$^{1,4}$.
These are generalizations of electromagnetic fields such that the
potential carries three antisymmetric
Lorentz indices and the field strength carries four. In four dimensions
these fields are non-dynamical - the field equations have only constant
solutions. If these field strength values can settle down at any value, they
can provide the random component.

Such four-forms do appear in string theory. In dimensions
greater than four they are dynamical with plane wave solutions. This
then gives a possible mechanism for producing them in a hot early universe
with energies above the compactification scale. When the
energy decreases through the scale of compactification, i.e.
when the world
becomes essentially four-dimensional, the four-form fields could
take on different values in spatially disconnected regions. With subsequent
inflation one of these regions could be the observable universe.

In string theory the values of the form fluxes are quantized, with the
field strength being related to the size of the compact dimensions$^4$.
The impact of this constraint appears to depend on the history.
In one extreme the moduli fields, governing the size and shape of the compact
dimension, reach the minimum of their zero-temperature potentials
first. Then the form fields
would only take on discrete values consistent with the quantization
constraint. For anthropic considerations to apply, the discrete
steps in the fields would need then need to be very tiny$^4$. However, if
the form fields are fixed first, before the moduli fields settle down to 
their absolute minima. they could potentially be at a continuous range of
values. In this case, the quantization condition provides 
the constraint on the minimization of the moduli fields.
The couplings between the form fields and the moduli imply that
random values of the form fields will generate random contributions
to vacuum selection and also imply random values of the
other physical parameters.

\section{Two Possible Tests}
\noindent
The consideration of such theories is new enough that we
do not fully understand their implications. Novertheless, I can offer two
modest possibilities that might be useful in revealing
such dynamics.

One signal is observational. If the physical parameters vary, the
cosmological constant will have the most dramatic variation. This is
because the cancellation required to obtain the observed value of
$\Lambda$ is enormous, and a tiny change in any parameter upsets
this cancellation. Therefore $\Lambda$ is most sensitive to
possible variation. If $\Lambda$ varies continuously it is possible that
it would have a small residual variation across our observed
universe. In effect, a variation from one side of the universe to the other
would signal that the parameter is not uniform and would hence favor this
type of theory. Unfortunately, we don't have firm predictions for the
size of the effect. A large amount of inflation could wash out a gradient in
the parameters leaving a small residual. Nevertheless, it is eventually
worth checking to see if there is any gradient in the cosmological
constant.

Another test could come once we have a fundamental theory that has been
shown to have multiple domains. This theory could not predict the specific
quark masses, as these would not be unique. However, it could predict the
distribution of quark masses. Empirically, the masses appear to
be distributed as if with a weight that is close to scale invariant,
$\rho (m) \sim 1/m$. This can serve as a test of the underlying theory.

\nonumsection{References}

\end{document}